\documentclass[12pt]{article}
\usepackage[cp1251]{inputenc}
\usepackage[english]{babel}
\date{}

\begin{document}

\title{Magnetization in AIIIBV semiconductor heterostructures with the depletion layer of manganese}
\author{T.\,Charikova\/\thanks{charikova@imp.uran.ru, "Low Temperature Physics", 201….}, V.\,Okulov,  A.\,Gubkin, A.\,Lugovikh\\ 
{\normalsize\it Institute of Metal Physics RAS, Ekaterinburg, Russia}\\
K.\,Moiseev, V.\,Nevedomsky\\ {\normalsize\it Ioffe Institute, St Petersburg, Russia} \\
Yu.\,Kudriavstev, S.\,Gallardo\\ {\normalsize\it Dep. Ingenieria Electrica - SEES, Cinvestav- IPN, Mexico}\\
M.\,Lopez\\ {\normalsize\it Dep. Fisica, Cinvestav-IPN, Mexico} }\maketitle

\begin{abstract}

The magnetic moment and magnetization in GaAs/Ga$_{0.84}$In$_{0.16}$As/GaAs  heterostructures with Mn deluted in GaAs cover layers  and with atomically controlled Mn $\delta$-layer thicknesses near GaInAs-quantum well ($\sim$3 nm) in temperature range T=(1.8-300)K in magnetic field up to 50 kOe have been investigated. The mass magnetization all of the samples of GaAs/Ga$_{0.84}$In$_{0.16}$As/GaAs with Mn  increases with the increasing of the magnetic field that pointed out on the presence of low-dimensional  ferromagnetism in the manganese depletion layer of GaAs based structures. It has been estimated the manganese content threshold at which the ferromagnetic ordering was found.


\end{abstract}
\section{Introduction}
Diluted magnetic semiconductors (DMSs) on III-V based materials have of greate interest both for the researchers in condensed matter physics and for the technologists  because of the semiconductivity-magnetism coexistence and  of the low equilibrium solubility of the transition metal \cite{Ohno1996}. DMS heterostructures are attracting attention due to the possibility of information management using  both the charge and the spin of carriers \cite{Nazmul2005,Dietl2007}.
Advances in epitaxial growth technology, such as molecular beam epitaxy, have made it possible to grow a variety of semiconductor heterostructures with atomically controlled layer thicknesses and abrupt doping profiles, in which the wave function of carriers within the artificially designed potentials may be controlled. Unlike the random alloy system, $\delta$-layer of Mn in GaAs provides the doping profile along the growth direction with the inherent advantages of  $\delta$-doping that it yelds  higher dopant carrier concentration \cite{Parfeniev2009}.
There is a large scatter in conductivities and Curie temperatures obtained by different groups in Ga(Mn)As materials \cite{Jungwirth2006}. These differences depend on the details of the growth and post-growth annealing indicating the Mn quantity and the presence of compensating defects.

Unlike the random alloy system, Mn $\delta$ -layer in GaAs provides the doping profile along the growth direction which can be approximated by Dirac's  $\delta$-function. Inherent advantages of  $\delta$-doping has locally higher dopant concentration and higher carrier concentration. In this paper we have presented  the results of the magnetization in heterostructures  GaAs/Ga$_{0.84}$In$_{0.16}$As/GaAs with Mn deluted in GaAs cover layers  and  by secondary ion mass spectroscopy (SIMS) depth profile analyses  Mn $\delta$-layer near GaInAs-quantum well.
 
\section{Experimental details}

The GaAs/InGaAs/GaAs quantum well (QW) heterostructures were grown on GaAs(001) substrate by MBE method in Riber C21 chamber at temperature range of 500-600 $^0$C \cite{Yee-Rendon2004}. Epitaxial deposition was carried out under average rate of 0.5  m/hour. The 300 nm-thick GaAs buffer layer as a first barrier of the QW was obtained at higher temperature (600 $^0$C) than the layer of the GaInAs ternary solution and the 5 nm-thick GaAs second barrier (500 $^0$C). The temperature decreasing was done to prevent diffusion of indium from QW to the GaAs barrier layers. Then, the temperature of a substrate was moving down less than 300 $^0$C to achieve optimal conditions for an atomic Mn layer deposition which was covered by an additional (cap) GaAs layer as thick as 50 nm. 
Profile analysis of the heterostructures containing the Mn layer was performed by Secondary Ion Mass Spectrometry (SIMS) method. Measurements were carried out using ion-microprobe ims-6f (Cameca, France). We have applied the de-convolution procedure to experimental SIMS depth profiles for the obtained InGaAs/GaAs single quantum well reported elsewhere \cite{Kudriavtsev2014}. The concentration of the element of interest was re-calculated by using the Relative Sensitivity Factors (RSFs) defined earlier by SIMS profiling of implanted standards. The thicknesses of the QWs and GaAs cover layers were examined by cross-sectional TEM. The TEM data were used as references in examining the de-convolution process we suggested and its application to experimental SIMS depth profiles.

The magnetic field dependencies of the magnetic moment m(H) and mass magnetization for GaAs/Ga$_{0.84}$In$_{0.16}$As/GaAs/Ga(Mn)As and GaAs/Ga$_{0.84}$In$_{0.16}$As/GaAs/$\delta$-Mn/GaAs heterostructures have measured using SQUID-magnetometer MPMS of Quantum Design in the temperature range T=(1.8-300)K and in magnetic field up to 50 kOe (Institute of Metal Physics RAS).

\section{Experimental results and discussion}

The magnetic field dependencies of the mass magnetization $\sigma$(H) for the structure GaAs/Ga$_{0.84}$In$_{0.16}$As/GaAs/Ga(Mn)As at low temperatures is presented in Fig.1. The measurements were done in two magnetic field orientations parallel, along with the substrate plane, and perpendicular, when the magnetic field was applied along the growing direction.  For comparison, GaAs substrate exhibited a large diamagnetic response in the parallel orientation and its mass magnetization reached $\sigma$ = -2.3*10$^{-3}$ emu/g in the field H = 10 kOe . As one can see in the figure that the field dependence of the magnetization does not exists in the case of the perpendicular magnetic field. So there is the anisotropy of the magnetization field dependence along  and perpendicular to the magnetic easy axis.

To extract the contribution of the Mn magnetic moments to the total magnetizationin the heterostructures with the barrier based on the Ga(Mn)As diluted compound, we have made several steps to distinguish the large common diamagnetic signal of the substrate (the thickness $d_s$= 0.5 mm) from the weak signal of the heterostructure ($d_h$=350 nm). The epitaxial wafers with the heterostructure on the GaAs substrate were processed after MBE growth to remove atomic indium. Then, the magnetic moment and the mass of the samples and the substrate were measured. Next, the contribution of the substrate from a sample mass magnetization was subtracted. This procedure is also required in order to exclude the possible contribution to the magnetization of random magnetic impurities such as iron.

The extracted dependencies of the mass magnetization for the samples either with diluted GaMnAs and the GaAs barrier locally doped with Mn $\delta$-layer are presented in Figure 2. It is obvious, when the magnetic field H is applied in the direction of the easy magnetic axis, magnetization  $\sigma$(H) for both samples increases with the increasing in the magnetic field. For the Ga(Mn)As sample the saturation of the magnetization was found out reaching value of   $\sigma$ $\simeq$ 6.4*10$^{-4}$emu/g at H = 10 kOe at T = 5K. The paramagnetic response was observed at T = 300K too (Fig.2, insert). Only weak signal of the mass magnetization was observed for the heterostructure with the Mn  $\delta$-layer (0.5 ML) placed into GaAs cap layer and separated from the InGaAs quantum well by the GaAs spacer layer as thick as 3 nm. Linear-like dependence without saturation was reaching $\sigma$ = 1.4*10$^{-4}$emu/g at H = 10 kOe. We suppose that the ferromagnetism in GaAs heterostructure with Mn $\delta$-layer is quantitatively different even from the magnetic responce in the structures with the diluted Ga(Mn)As cover layer and differ qualitatively from their bulk counterparts \cite{Tripathi2014}. In the dilute limit the magnetization can be discribed by the Brillouin function \cite{Furdyna1988}:

\begin{equation}
    M=-xN_0 g_{(Mn)} \mu_B \langle S_x\rangle,
\end{equation}

where $\langle S_x\rangle$ is the average spin per Mn site, $N_0$ is a number of cations per unit volume, g - the g factor, $\mu_B$ is the Bohr magnetron and $x$ - molar fraction. Using the data of the mass-analyzer for GaAs heterostructures with diluted Mn $N_0$=$N_{Mn}$$\simeq$10$^{20}$ cm$^{-3}$, the value of saturation of magnetization at $T$=5K $\sigma$ $\simeq$ 6.4*10$^{-4}$emu/g, $g$=2 and $S$=$\frac{5}{2}$ for Mn$^{++}$ we have estimated the molar fraction of deluted Mn in the cover layers. This value $x$ $\simeq$0.002 corresponds the delute limit where Mn$^{++}$ spins are isolated. 

In magnetic field H = $\pm$1.5 kOe the mass magnetization shows a hysteresis loop in the structure with the diluted Ga(Mn)As layer (Fig.3) that indicates the ordered ferromagnetic structure.

\section{Conclusions}

It was experimentally found that ferromagnetism exist in the magnetic quasi two-dimensional GaAs-based heterostructures doped with digital layer of manganese at the temperature T = 5 K and T = 300 K even in the dilute limit. In the case of the heterostructures with the Mn $\delta$-layer (with concentration of 0.5-1 ML) the paramagnetic response that can be increased using further improvements in growth conditions of the $\delta$-layers due to layer thickness and Mn diffusion control to obtain more convincing evidence of the spin polarization. Work in this area should lead to sustainable results for their application in development of modern industry.
 
This work was done within RAS Program (project N 12-P-2-1018) with partial support of RFBR (grants N 12-02-00202 and N 12-02-00530). Authors from Cinvestav thank to SENER and CONACYT, both from Mexico for a financial support of this study, grant N 152244.

\newpage

Fig.~1. The magnetic field dependence of the mass magnetization $\sigma$ at T=5K for GaAs heterostructures with H parallel to the sample plane and perpendicular to it.

Fig.~2. The magnetic field dependence of the mass magnetization $\sigma$ at T=5K for GaAs heterostructures with diluted Mn and with Mn $\delta$-layer. $\sigma$(H) dependencies for GaAs heterostructures with diluted Mn at the temperatures T=5K and T=300K are presented insert.

Fig.~3. The magnetization hysteresis curve for GaAs/Ga(Mn)As. The lines are presented as a guide for the eyes and triangles are the first trip up, circles - down, squares - up once again.

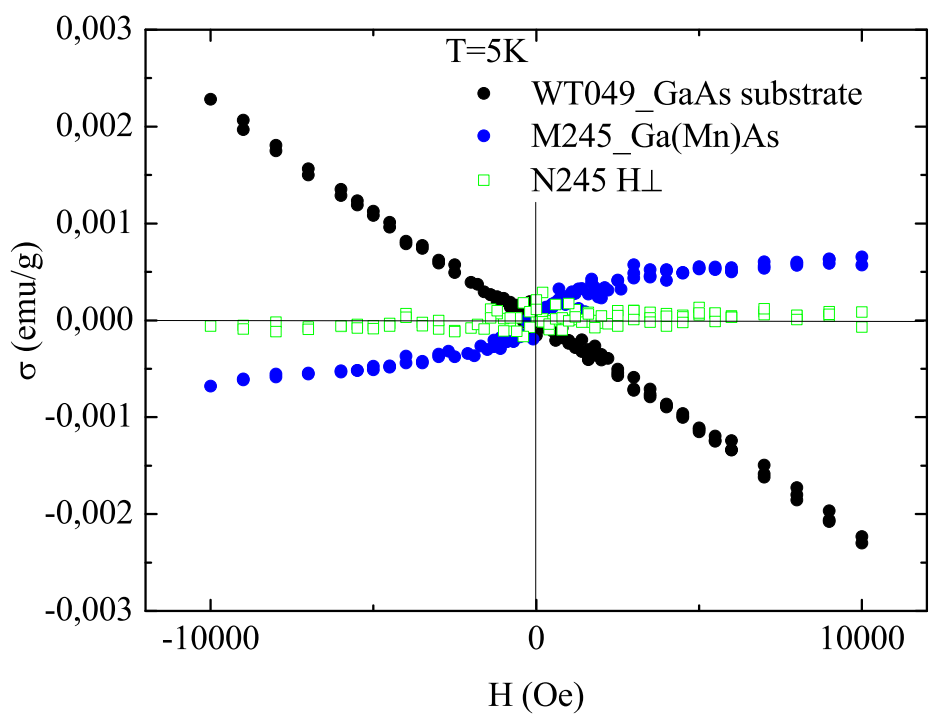
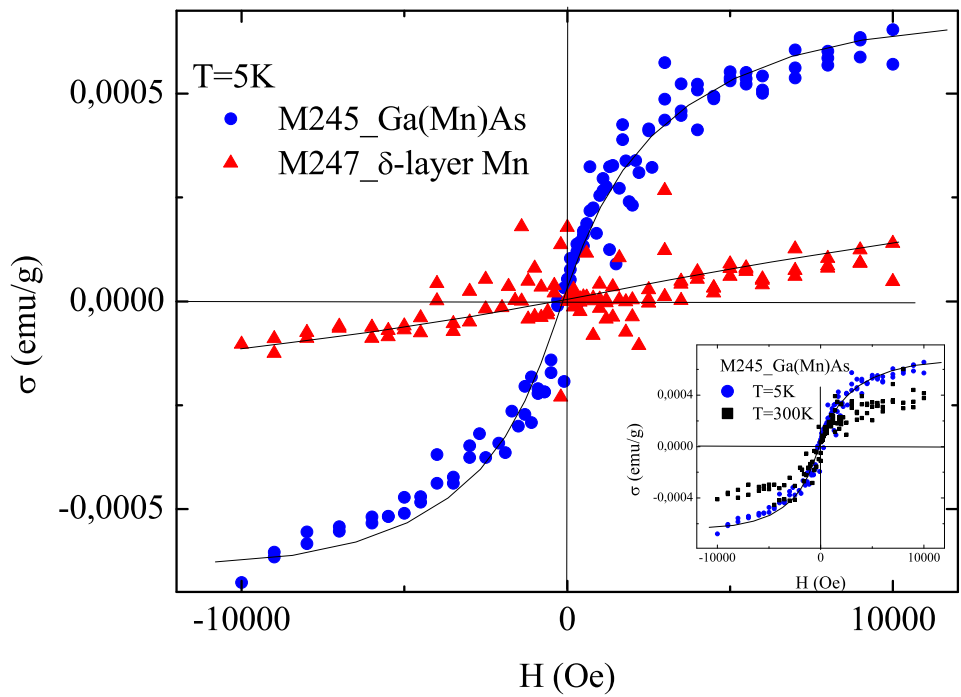
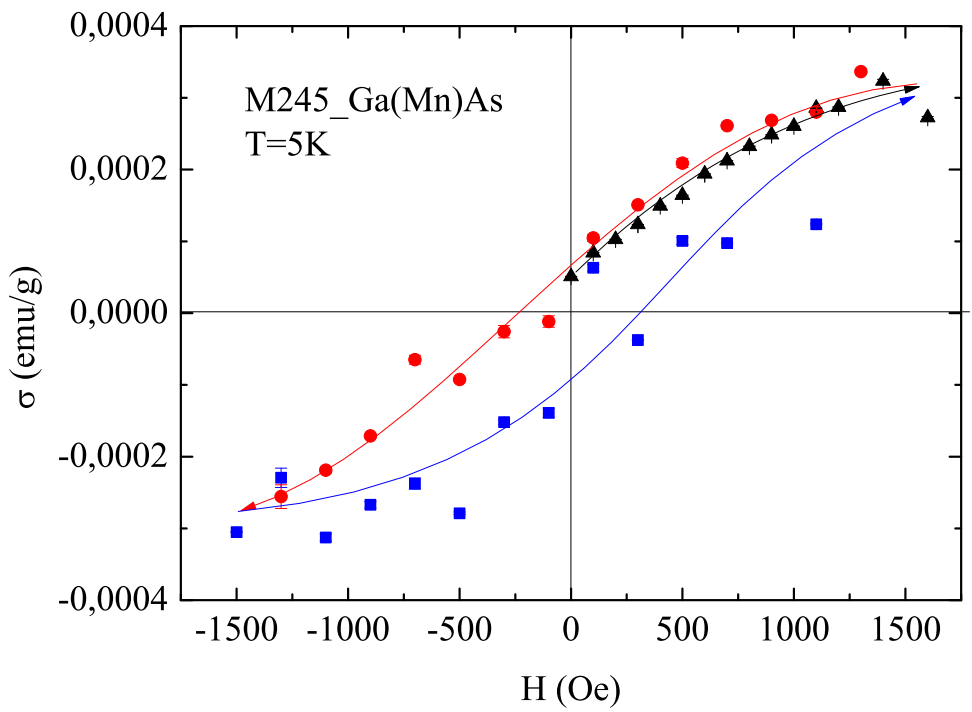
\end{document}